\documentclass[journal]{vgtc}                % final (journal style)
\ifpdf%                                % if we use pdflatex
  \pdfoutput=1\relax                   % create PDFs from pdfLaTeX
  \usepackage{graphicx}                % allow us to embed graphics files
  \DeclareGraphicsExtensions{.pdf,.png,.jpg,.jpeg} % for pdflatex we expect .pdf, .png, or .jpg files
\else%                                 % else we use pure latex
  \usepackage{graphicx}                % allow us to embed graphics files
  \DeclareGraphicsExtensions{.eps}     % for pure latex we expect eps files
\fi%

\graphicspath{{figures/}{pictures/}{images/}{./}} % where to search for the images

\usepackage{microtype}                 % use micro-typography (slightly more compact, better to read)
\PassOptionsToPackage{warn}{textcomp}  % to address font issues with \textrightarrow
\usepackage{textcomp}                  % use better special symbols
\usepackage{mathptmx}                  % use matching math font
\usepackage{times}                     % we use Times as the main font
\usepackage{cite}                      % needed to automatically sort the references
\usepackage{tabu}                      % only used for the table example
\usepackage{booktabs}                  % only used for the table example
\usepackage[hyphens]{url}
\usepackage{hyperref}

\onlineid{1039}

\vgtccategory{VIS 2020 VAST Papers}
\vgtcpapertype{application/design study}

\newcommand{\name}{{\textit{TaxThemis}}}

\newcommand{\yating}[1]{\textcolor{black}{#1}}
\newcommand{\jason}[1]{\textcolor{black}{#1}}

\title{{\name}: Interactive Mining and Exploration of\\ Suspicious Tax Evasion Groups}

\author{Yating Lin*, Kamkwai Wong*, Yong Wang, Rong Zhang, Bo Dong, Huamin Qu, Qinghua Zheng}
\authorfooter{
\item
Y. Lin and Q. Zheng are with the MOEKLINNS Lab, Xi'an Jiaotong University. Emails: linyating@stu.xjtu.edu.cn and qhzheng@mail.xjtu.edu.cn
\item
K. Wong, Y. Wang, R. Zhang, and H. Qu are with Hong Kong University of Science and Technology.
E-mails: \{kkwongar, ywangct, rzhangab\}@connect.ust.hk and huamin@cse.ust.hk
\item
B. Dong is with the National Engineering Lab of Big Data Analytics, Xi'an Jiaotong University.
E-mail: dong.bo@mail.xjtu.edu.cn
\item
*These authors contributed equally to this work.
}

\shortauthortitle{Biv \MakeLowercase{\textit{et al.}}: Global Illumination for Fun and Profit}
\abstract{
Tax evasion is a serious economic problem for many countries, as it can undermine the government's tax system and lead to an unfair business competition environment. 
Recent research has applied data analytics techniques to analyze and detect tax evasion behaviors of individual taxpayers. 
However, they have fail\jason{ed} to support the analysis and exploration of \jason{the} related party transaction tax evasion (RPTTE) behaviors (e.g., transfer pricing), where a group of taxpayers is involved. 
In this paper, we present {\name}, an interactive visual analytics system to help tax officers mine and explore suspicious tax evasion groups through analyzing heterogeneous tax\jason{-related} data. 
A taxpayer network is constructed \jason{and fused with the respective trade network} to detect suspicious \jason{RPTTE} groups. 
Rich visualizations are designed to facilitate the exploration and investigation of suspicious transactions between related taxpayers \jason{with profit and topological data analysis}. 
Specifically, \jason{we propose} a calendar heatmap with a \jason{carefully-designed} encoding scheme to intuitively show the evidence of transferring revenue through related party transactions.
We demonstrate the usefulness and effectiveness of {\name} through two case studies on real-world tax-related data and interviews with domain experts.
} % end of abstract


\CCScatlist{ % not used in journal version
 \CCScat{K.6.1}{Management of Computing and Information Systems}%
{Project and People Management}{Life Cycle};
 \CCScat{K.7.m}{The Computing Profession}{Miscellaneous}{Ethics}
}

\teaser{
  \centering
  \includegraphics[width=\linewidth]{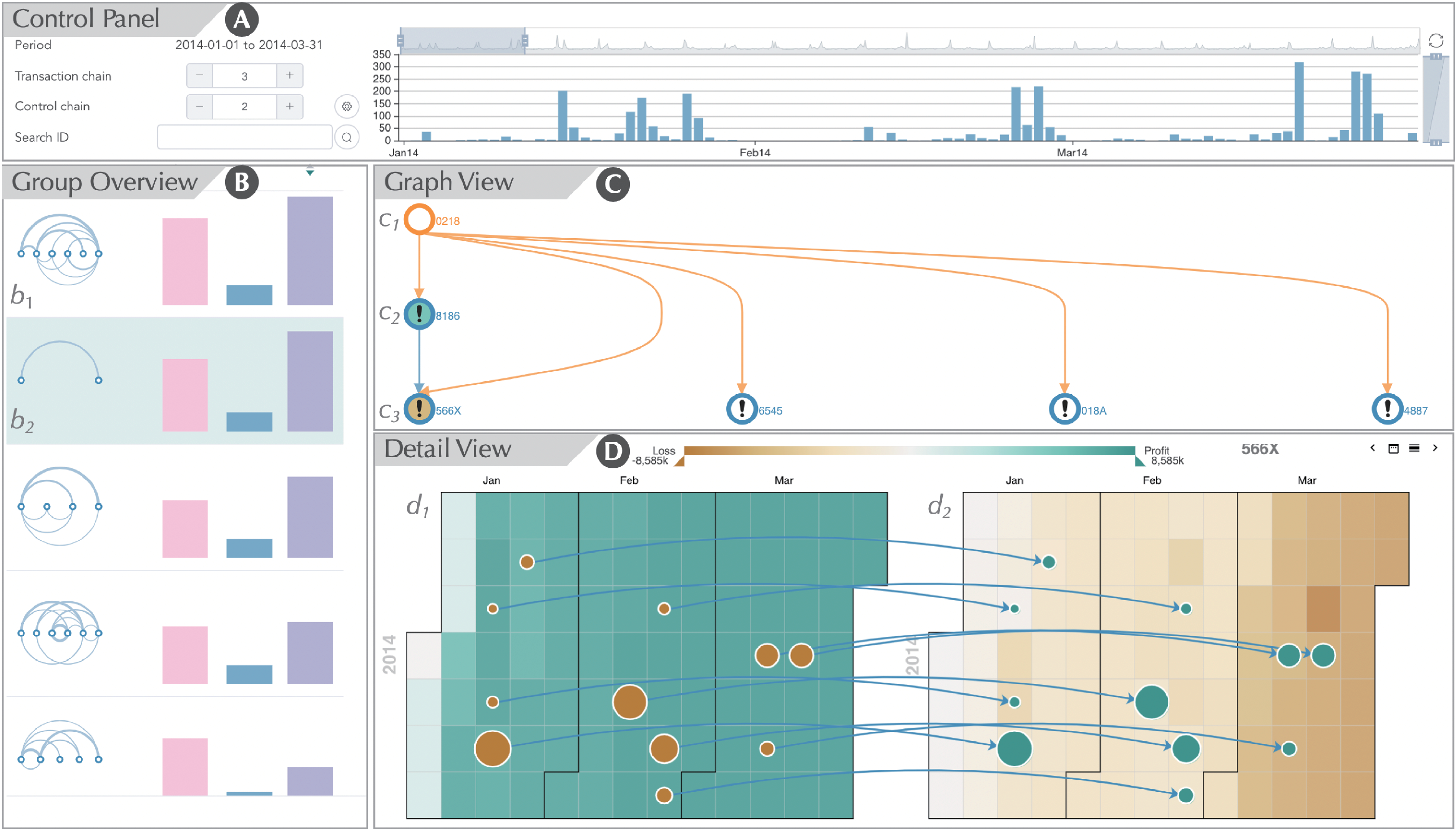}
  \caption{The system interface of {\name}: (A) The Control Panel consists of a bar chart to show the temporal summary of the daily related party's daily transaction amounts, together with a parameter selector \jason{to support network fusion through interactively setting the preferred period or relevant thresholds}. (B) The Group Overview visualizes \jason{the topology of related party transactions in each suspicious group, and ranks the groups by their group features. This helps users focus on the most suspicious groups.} (C) The Graph View shows the hierarchical investment \jason{relationships and related party transactions within the selected suspicious} group. \jason{It supports group assessment in revealing the common beneficial owners and former tax evaders}. (D) The Detail View presents \jason{the profit status of the traders who conducted the selected related party transactions}, revealing their suspicious behavior of transferring revenue transfers.}
  \label{fig:system_overview}
}

\vgtcinsertpkg

\begin{document}
\maketitle


% \firstsection{Introduction}

% \maketitle

\section{Introduction}
\label{sec-intro}

\maketitle

% 1. Introduce the background and importance of detecting suspicious tax evasion group. Specifically, we need to emphasize the importance of detecting suspicious tax evasion group. You can refer to the introduction part of the 5 papers in the ``related work/tax'' folder. Then, we briefly discuss the data we have, which makes the data-driven detection of suspicious tax evasion group possible. 

%%% the severe tax evasion problem and its influence %%%
% The definition of tax evasion
Tax evasion is a financial crime against the tax system in which taxpayers deliberately report false financial positions to deny their tax obligations.
It has become a severe economic problem for many countries resulting in a significant \textit{tax gap}, i.e., the difference between the tax revenue the government is entitled to and the actual revenue collected.
% The monetary impact of tax gap, giving the ratio to GDP which more people understands
For example, the predicted tax gap for 28 European Union countries \yating{was, on average, \$51.9 billion}, \yating{nearly} 7.7\% of their 2015 GDP~\cite{global_tax_gap_2018}. 
In the United States, the estimated tax gap was at \$441 billion for 2011-2013\jason{, or about 2.7\% of their GDP}~\cite{federal_2019}.
\jason{In Australia, the estimated tax gap was at \$21.9 billion in 2015, roughly 2.2\% of its GDP~\cite{ato_2020}.}
% Other impacts to show the importance
Aside from less revenue for government budgets, tax evasion has also led to an unfair business competition environment, especially for compliant taxpayers whose operating costs are consequently higher than their non-compliant competitors. Such a phenomenon would eventually incur more tax non-compliance to challenge the tax system~\cite{oecd_2017}.

% Reassure why tax evasion detection is important
Tax authorities have expended tremendous efforts in trying to reduce the tax gap and curtail tax evasion behaviors.
Many computer-aided case selection approaches have been widely used to detect individual tax evasion behaviors, such as account manipulation~\cite{liu_application_2010, wu_using_2012, hsu2015data, fraudster_brasil_2015, kim_detecting_2016} and false invoices~\cite{gonzalez2013characterization}.
However, these approaches mainly focus on detecting individual tax evaders by designing a set of financial indicators and further training a classifier model to classify the taxpayer entities. 
They are not able to detect \textit{Related Party Transaction-based Tax Evasion(RPTTE)} behaviors, where groups of related taxpayers are involved in illegally redistributing their profits and losses to reduce the overall tax burden~\cite{UN_manual_2017, ruan_identifying_2019}.
The RPTTE behaviors are often conducted through related party transactions and transfer pricing and have become a growing global trend in tax evasion~\cite{klassen2017transfer, ferrantino2012evasion, UK_transfer_pricing_2017}.
Thus, the effective mining and exploration of tax evasion groups are of significant importance in reducing the tax gap.

However, it is challenging to analyze and explore tax evasion groups effectively. 
First, the detection of tax evasion groups depends on the analysis of the topological interest relationship between different taxpayers and their various tax-related attributes, making it quite complex to explore tax evasion groups.
Second, the ambiguity in the accounting principle has resulted in a time-consuming audit procedure for tax officers to check manually if a suspicious group conducts tax evasion. Even the recent state-of-the-art approaches for automated tax evasion group detection~\cite{tselykh_attributed_2016, tian_mining_2016} can result in a high false-positive rate and an overwhelming number of suspicious cases.
Third, correlating a large amount of financial data and related party transactions and their analysis is time-consuming, complicating the exploration of suspicious patterns of tax evasion groups.

%% 3. Introduce our work and briefly discuss how our approach will handle the challenges discussed above.
To handle the above challenges, we attempt to employ visual analytics to assist tax officers with the mining and exploration of suspicious tax evasion groups. We have worked closely with a provincial tax administration office in China for two years to understand the detailed requirements of tax evasion detection and exploration. We investigated the detailed tax-related data, including taxpayer profile information, audit, trade, and investment records. Our collaborators in the tax administration office provided all the data used in this paper after conducting data masking on the original data. We aim to tightly integrate the domain knowledge of tax officers with the computational power of machines and the expressiveness of visual analytics. We present {\name}, an interactive visual analytic system, to help tax officers explore and identify suspicious tax evasion groups. We first propose a network fusion algorithm adapted from a graph-based tax evasion group detection approach~\cite{tian_mining_2016} to identify any suspicious RPTTE groups, where the formulation of the taxpayer network was modified according to the requirements posed by our collaborators. Furthermore, we propose a set of coupled visual designs to assist tax officers in the exploration of suspicious RPTTE groups. 

As shown in Fig.~\ref{fig:system_overview}, {\name} consists of four major views: the Control Panel, the Group Overview, the Graph View, and the Detail View. 
\jason{
The Control Panel consists of a bar chart showing the temporal summary of the daily related party transaction amount, and a parameter selector to support network fusion through interactively setting the preferred period or relevant thresholds. 
The Group Overview shows a list of suspicious RPTTE groups, in which each row represents a suspicious RPTTE group and consists of an arc-diagram based glyph and a bar chart to help users focus on the most suspicious groups. 
The Graph View shows the investment and trading relationships within the selected suspicious group.
The Detail View shows the profit status of the traders who conducted the related party transaction for profit analysis.}
We have applied this system to the real tax dataset and demonstrated its effectiveness through case studies and expert interviews.
Our contributions can be summarized as follows:

% 4. Briefly discuss the evaluations (case studies and expert interviews) and list the major contributions of our approach:
\begin{itemize}
\item \yating{We formulate the design requirements for interactive mining and visual exploration of tax evasion groups, together with two domain experts, that is, an experienced tax officer from a provincial tax administration office and a tax researcher from a university in China.}

\item \yating{We present an interactive visual analytics system, {\name}, to enable the quick exploration and inspection of tax evasion groups with topological and profit analysis. The proposed intuitive and carefully-designed visualizations provide an overview and detailed explorations of tax evasion groups. To the best of our knowledge, {\name} is the first visual analytics system that is designed for detecting and exploring tax evasion groups by analyzing related party transaction-based tax evasion (RPTTE) behaviors.}

\item We conducted two case studies with a real tax dataset and well-structured expert interviews with two domain experts, which demonstrate the effectiveness and usefulness of {\name}.
\end{itemize}

\section{Related Work}
\label{sec-relatedwork}
In this section, we summarize the techniques that are most related to our work, including tax evasion detection methods and visualization techniques for financial data.

\subsection{Tax Evasion Detection}
\label{subsec-relatedwork}
The majority of the literature in computer-based tax inspection falls into two categories: machine learning methods and network-based methods.

\textbf{Machine Learning Methods.} These methods mainly focus on proposing models to calculate the suspicion scores of taxpayers’ fraud intentions~\cite{gonzalez2013characterization,Rahimikia_iran_2017,kim_detecting_2016,hoglund_tax_2017}. For example, the Chilean administration used decision trees, neural networks, and Bayesian networks to calculate the suspicion scores of taxpayers based on the features extracted from the information of taxpayers and the related invoice data~\cite{gonzalez2013characterization}.
The Iranian tax administration proposed a hybrid model to detect corporate tax evasion by using the multi-layer perceptron neural network, support vector machine, and logistic regression classification models in combination with the harmony search optimization algorithm~\cite{Rahimikia_iran_2017}.
Kim et al. developed a multi-class predictive tool to detect fraudulent financial misstatements~\cite{kim_detecting_2016}. 
Höglund developed a genetic algorithm-based decision support tool for predicting tax payment defaults~\cite{hoglund_tax_2017}. 

Although the machine learning-based methods can classify taxpayers with suspicion scores calculated by pre-trained models, they are not explainable to tax inspectors and are highly reliant on training data. It is therefore difficult to verify the methods' efficiency since they have a high false positive rate.

\textbf{Network-based Methods.} These methods aim to explore specific patterns of tax evasion based on a social network of taxpayers, which is the so-called taxpayer interest interacted network(TPIIN)~\cite{tian_mining_2016,ruan_identifying_2019,tselykh_attributed_2016}. Tian et al. first proposed a color-based model to represent the type of taxpayer and their relationships. They summarized tax evasion patterns based on audit cases and presented a pattern matching algorithm to explore suspicious tax evasion groups in terms of graph theory~\cite{tian_mining_2016}.
To conduct comprehensive research on the affiliated-transaction-based tax evasion group, Ruan et al. proposed a hybrid method for identifying suspicious groups via three steps: tax rate differential detection, topological pattern matching, and tax burden abnormality identification~\cite{ruan_identifying_2019}.
Other work by Tselykh et al. presented an attributed-graph-based approach to detect suspicious fraudulent transfer pricing entities with clustering and rule induction techniques~\cite{tselykh_attributed_2016}.

Research conducted by the network-based method has effectively detected typical tax evasion groups with specific expert-summarized patterns. However, the algorithms generate a large number of suspicious groups that contain a lot of false positives and need to be identified with extra effort. 

\subsection{Financial Data Visualization}
A number of visualization techniques have been developed for financial data, \yating{in view of its complex data structure and significant impact in application fields. We have summarized related financial data visualization works into three categories: time-series data visualization, multivariate graph visualization, and financial crime visualization.}

\textbf{Time-series Data Visualization.}
\yating{Temporal data is prevalent in many problem domains. 
The visualization community has developed numerous visual techniques and have applied them in analyzing time-series data in different fields, which facilitate the detection, exploration, and prediction of several specific tasks. 
For example, Guo et al. proposed a visualization system with a thread-based design to support progression stage identification in time-series events~\cite{guo_visual_2019}.
Malik et al. proposed a visualization framework for spatio-temporal correlations analysis~\cite{malik_correlation}.}
With the popularity of digital transactions, a mass of time-series trade data has been collected by financial institutions, which promote the adoption of visual analytics approaches to analyze transaction events.
Previous work on visualizing time-series trade data aims at exploring correlations between trade goods, network patterns, and time-series signatures.
Hong Wang et al. proposed a spatio-temporal trade network visualization to explore the relationship between global trade networks and regional instability~\cite{wang_visual_2019}.
Xie et al. also introduced a visualization system called “VAET” to detect salient transactions from large e-transaction time-series~\cite{xie_vaet_2014}.
Work by Yue et al. proposed a novel dynamic timeline visualization to explore the evolutionary transaction patterns of Bitcoin exchanges from two perspectives~\cite{yue_bitextract_2019}. 
Instead of focusing on \yating{the process and the correlation}, this paper concentrates on the analysis of \yating{the continuous profit time sequences of} suspect traders in a tax evasion group.

\yating{\textbf{Multivariate Graph Visualization.}
Multivariate graph visualization is widely used in topological data analysis as it is intuitive and interactive. To facilitate global exploration, Elzen et al. proposed a large-scale multivariate network visualization method with rich interactions to support selections and aggregations~\cite{van_den_elzen_multivariate_2014}.
Focusing on exploring the local subgraphs, Pienta et al. presented common visual encoding schemes to embed the features of nodes and links~\cite{pienta_vigor_2018}.
For intuitively displaying the network topology together with the correlation analysis of multiple entity attributes, Nobre et al. proposed the tree+table multivariate graph visualization technique~\cite{nobre_juniper_2018}.
Different from previous works, our works focus on analyzing any suspicious phenomenon by showing its topology and suspicion attributes of entity to help locate the most suspicious links, then further show the context of the link to facilitate its identification.}

\textbf{Financial Crime Visualization.} The visual analytic technique has proved its importance in detecting financial fraud~\cite{dilla_data_2015, kielman_foundations_2009, leite_visual_2018}.
An early work of Kirkland et al. adopted artificial intelligence, visualization, pattern recognition, and data mining to support fraud detection analysis~\cite{Rahimikia_iran_2017}.
Huang et al. proposed a VA system, and use 3D treemaps to monitor the real-time stock market performance and identify a particular stock that produced an unusual trading pattern~\cite{huang_visualization_2009}.
Work by Chang et al. proposed WireVis, using a keyword graph, the strings, and bead visualization to allows the analysts to see the complete relationship between accounts, keywords, time, and patterns of activity in transaction data to support fraud detection~\cite{chang_wirevis_2007}.
Didimo et al. presented VISFAN, a visual analytics system that uses social network analysis combined with a clustering algorithm to support analysts with effective tools to discover financial crimes, like money laundering and fraud in financial activity networks~\cite{didimo_advanced_2011}. 
Work by Leite et al. proposed EVA, which offers a scoring mechanism with interactive visual analytics facilities to perform fraud validation~\cite{leite_eva_2018}.
Recently introduced are a few analytic systems for tax fraud detection. 
The Italian Revenue Agency proposed a decision support system for detecting specific patterns of tax evasion groups~\cite{didimo_visual_2018}. 
They later applied a system based on pattern matching and risk information diffusion to support public officers in identifying crafty taxpayers~\cite{didimo_combining_2020}.

Our work differs from others in that we mainly focus on the task of exploring the RPTTE group. We have come up with several coordinated visualizations with \yating{carefully-designed} visual encoding schemes to support interactive mining and the exploration of suspicious groups. 
% briefly introduce the background with data description and task&requirement analysis
\section{Background and Data Description}
\label{sec-background}

\yating{As introduced in Section~\ref{sec-intro}, we aim at detecting tax evasion groups to reduce the tax gap. Some core concepts regarding tax evasion are clarified below to help comprehend the background.
A taxpayer is another taxpayer's \textit{related party} if they share a common beneficial owner or are connected in an interest relationship other than investment (e.g., kinship).
\textit{Related party transaction} refers to the transferal of assets, goods and services between related parties.
Conducting a related party transaction is legal as long as the traders satisfy their tax obligations~\cite{oecd_oecd_2017}.
However, when such behavior is conducted to redistribute their profits and losses to reduce the overall tax burden, it is criminal behavior and is referred to as \textit{Related Party Transaction-based Tax Evasion(RPTTE)}~\cite{UN_manual_2017, ruan_identifying_2019}. 
The set of taxpayers who perform RPTTE and the corresponding investors form a \textit{RPTTE group}. 
In this paper, we focus on the detection and exploration of RPTTE groups.}

\yating{We analyze the raw tax-related data to ensure it supports our analytic approach.}
We have worked closely with two domain experts, one of whom is a tax officer from a provincial tax administration and one is a tax evasion detection researcher from a cooperative research group.
They have guided us through the tax-related data that they possess and use in their current tax inspection routine.
\yating{One main duty of the tax administration officers is to find related party transactions and search for evidence, which is often done by carefully examining the tax-related data and monitoring certain confidential financial indicators.} 
Tax-related data includes taxpayer and investor profiles, invoice information, and audit records. 
We provide a brief introduction to the three types of tax-related data used for tax evasion detection and analysis.

\textbf{Taxpayer and Investor Profiles:} A taxpayer is a person or organization (such as a company) who needs to pay taxes to the government.
\yating{An investor is a person or organization that buys shares in taxpayers and holds voting rights.}
Our collaborator offered us the profile information of each taxpayer and the corresponding investors. 
The taxpayer profile describes the business nature of taxpayers, such as industry, major merchandise, ownership type, and so on. The corresponding investor information includes investor entity type, investment amount, and share ratio.
There are over 4 million taxpayers and 0.9 million investors in the entire dataset provided by our collaborator. 
\jason{This information helps us understand the topology of investment relationships of all taxpayers.}
  
\textbf{Invoice Information:}
\yating{The invoice information is collected to record the details of each transaction between taxpayers.}
We obtained 14 million \textit{Value-Added-Tax (VAT)} invoices in Shaanxi Province and the time range of the invoices is from Jan 1, 2014 to Dec 31, 2015. 
Each invoice record consists of five attributes: date, seller, buyer, VAT tax amount, and the transaction amount. 
The buyer and seller in an invoice record are two taxpayers, of whom, at least one of them, are registered in Shaanxi Province. 
\jason{The invoices explain the trading relationships among the taxpayers and show the cash flow of the taxpayers regarding transactions.}

\textbf{Audit Records:} Audit records refer to the results after a tax administration officer conducts an official examination on a taxpayer's financial account.
Our collaborators offered us historical audit records to help analyze each case and develop our system. 
Each auditing record consists of audit date, violation type, case description, action taken, and tax payable.
\jason{Together with the taxpayer profile, we can trace the tax evasion history of taxpayers.}

% \section{Design Considerations}
\section{Requirement Analysis}
\label{sec-requirment}

To better understand user requirements when users analyze and explore tax evasion groups, we have been working closely with two domain experts \textbf{E1} and \textbf{E2} over the past two years.
We recruited experts from our collaborating organization (a provincial tax administration and a research group).
Expert \textbf{E1} is a tax officer who has worked in the tax office for over 20 years and is proficient in tax risk management and audit procedures in China.
His job duties are to identify high-risk companies with the risk management system, interview suspicious companies and investigate their reported financial statements.
Expert \textbf{E2} is a tax researcher who has been working on tax evasion research projects over the past five years and has cooperated closely with at least five provincial tax offices in China. 
His research interests focus on developing fast tax evasion group detection algorithms to improve the efficiency of the tax inspection.

We conducted a series of structured face-to-face interviews and remote meetings with the domain experts. 
\yating{We are concerned with three aspects of tax evasion inspection: 
1) What are the common practices and their procedures for detecting and analyzing tax evasion groups? 
2) What are the major challenges and limitations of the current methods for finding and exploring tax evading groups? 
3) What kind of design requirements and tasks do they want to achieve?}

\jason{Through the interview sessions, the experts clarified the current audit procedure which mainly consists of three steps.
First, the tax evasion group detection conventionally relies on random selection and confidential financial indicators calculated from taxpayers' period-end financial positions. 
Second, the tax administration officers summon and interview the accounting representatives who prepare and disclose the financial statements for the suspected taxpayers.
To facilitate the fact-finding process, the officers may conduct field work and request additional proper business records.
Third, the transactions and information gathered are examined to discover supporting evidence and circumstantial evidence.}
Such an operation for tax evasion group detection and exploration involves a massive amount of manual checking of the raw tax related data, which is tedious and time consuming.

\yating{To harness the computational power of machine, some automated tax evasion group detection algorithms have already been proposed and used for detecting suspicious tax evasion groups~\cite{tian_mining_2016,ruan_identifying_2019}.
However, these automated algorithms, which focus on the topological patterns of tax evasion groups instead of their tax evasion behaviors, can produce a lot of false positives which are costly to verify. 
They are also unable to provide the suspicious invoices for tax administration officers to make audit decisions upon or prioritize, owing to their limited investigative resources.
Another key design goal is how to handle these issues for {\name}.
With these limitations in mind, we summarized the feedback from experts and further compiled a list of design requirements to guide the designs of {\name}.}

\begin{itemize}
\item[\textbf{R1}]\textbf{Enable interactive configurations for suspicious RPTTE groups detection.}
\yating{The automated algorithms significantly improve the efficiency of the tax inspection procedure when they can successfully identify the most relevant suspicious RPTTE groups.
As mentioned by \textbf{E2}, different users may want to explore RPTTE groups that satisfy specific conditions and would like to configure different parameters to extend or narrow down the scope of suspicious groups by filtering taxpayers not tightly connected to the related party transactions.}
For example, different users may have an interest in detecting tax evasion groups from different periods, and exploring tax evasion groups with a specific relevance or complexity.

\item[\textbf{R2}] \yating{\textbf{Rank suspicious tax evasion groups with multiple criteria.} Given the vast number of suspicious groups, it is critical to help users quickly locate groups with the highest risk level. The system should support the sorting of the groups based on multiple criteria which reflects the suspicion in different dimensions. For example, according to \textbf{E1}'s audit experience, one of the major suspicion criteria is the existence of historical tax evasion records because those groups are likely to commit tax evasion again. 
}

\item[\textbf{R3}]\yating{\textbf{Support the interactive exploration of the common beneficial owners of taxpayers who conduct the related party transactions and their attributes.}
As at least one common beneficial owner will benefit from RPTTE behaviors, exploring the investment and trading relationships helps users to understand how the tax evasion scheme works among taxpayers. 
In addition, the attributes of taxpayers such as historical tax evasion records provide the tax administration officers a context for suspicion and risk evaluation.
Therefore, the experts require that all taxpayer with a common beneficial owner should also be clearly visualized, facilitating the deep exploration and inspection of any highly suspicious related party transactions.}
\item[\textbf{R4}]\yating{\textbf{Provide convenient profit analysis of taxpayers that conducted related party transactions.}
% \yong{What is the motivation or reasons of analyzing profits here? Please further enhance it.}
To facilitate the tax inspection process, the users need to know how the taxpayers redistribute their profits through RPTTE behaviors. 
It is also important to present the related party transaction as evidence for users to quickly make audit decisions.
Both \textbf{E1} and \textbf{E2} agree that the evaluation of suspicious related party transactions is challenging because tax evaders try their best to disguise their transactions as legal ones. 
However, the intent of such transactions must be reflected in reported profits, which leads to a lower tax burden.
Therefore, profit analysis can act as the critical context to help users in decision-making.
The visualization should display the profit status of the taxpayers, as the analysis of profit variations can reveal whether the related party transaction behaviors affect the overall tax burden.}
%  \yong{We may add a requirement called intuitive visualization design.}
\end{itemize}
\begin{figure*}[!ht]
    \setlength{\belowcaptionskip}{-0.3cm}
	\centering
	\includegraphics[width=1\textwidth]{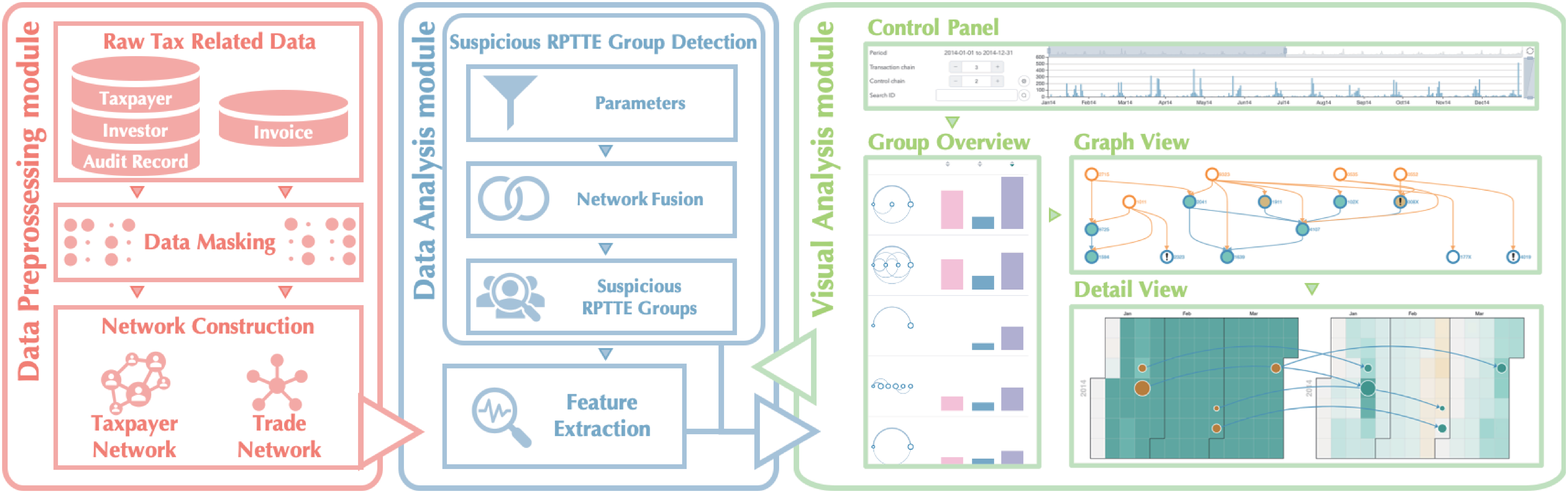}
    \caption{System architecture and data processing pipeline. {\name} consists of three major modules: the Data Preprocessing module, the Data Analysis module, and the Visual Analysis module. The Data Preprocessing module preprocesses the raw tax-related data with the data masking techniques and constructs the taxpayer network and the trade network. The Data Analysis module fuses the two networks based on the parameters and extracts features of each group to help further ranking and analysis. The Visual Analysis module supports interactively mining and exploration of suspicious RPTTE groups through four coordinated views.}
    \label{system_architecture}
\end{figure*}

\section{System Design}
% brief introduction of function and method
\jason{We propose an interactive tax inspection system, {\name}, to help tax administration officers explore and identify suspicious RPTTE groups.}
By following the design requirements discussed in Section~\ref{sec-background}, {\name} first discovers groups of taxpayers who conducted related party transactions within a specific period and then examines their group characteristics as well as financial performances.
Our system integrates the traditional data mining algorithm with visual analytics techniques. 
% Introduce the main modules of system architecture and data processing pipeline
As shown in Fig.~\ref{system_architecture}, {\name} consists of three major modules: the Data Preprocessing module, the Data Analysis module, \jason{and }the Visual Analysis module. 

% briefly introduce the function and purpose of each module
The \textbf{Data Preprocessing} module performs both data masking and \yating{network construction}.
Data masking aims to avoid revealing the identity of taxpayers and guarantees privacy protection for taxpayers, which is critically important in such sensitive data exploration.
\jason{Network construction refers to the extraction of the investment and trading relationship from the masked data.}
Inspired by the widely-used graph mining algorithm~\cite{tian_mining_2016}, we construct \jason{a taxpayer network and a trade network to model the two relationships.}
%\yong{Yating, pls clearly clarify which algorithm you are using by cititing the paper.}

\jason{The \textbf{Data Analysis} module fuses the taxpayer network and the trade network based on the parameters used to detect suspicious RPTTE groups.}
To help different users perform RPTTE group detection according to their exploration goals, we enable users to determine the extent of the suspicious RPTTE groups interactively. 
To facilitate the convenient exploration of these groups, we further extract tax evasion related features from them. These features can either characterize the RPTTE groups or delineate the related party transactions).

The \textbf{Visual Analysis} module presents the suspicious RPTTE groups together with the corresponding contexts and detailed evidence \jason{through intuitive visualizations. It facilitates a convenient exploration and in-depth examination of RPTTE groups with these linked visualization views.}
% Briefly introduce the implementation way of all modules
All these modules are built and integrated as a web-based system, where the backend is developed with Python and Flask to support data preprocessing and data analysis. The frontend is implemented by JavaScript, Vue, and D3 to build visualization views and support interaction.

\subsection{Data Preprocessing Module}
This module turns the raw data from our collaborators into a \jason{taxpayer network and a trade network}.

\textbf{Data Masking.}
Since the profile information of taxpayers and investors must be kept confidential, we preprocess the raw data with data masking. 
For identifiable taxpayer information, we add scrambled characters and \jason{masked specific fields of taxpayers' information so they are not easily identifiable.}
For the invoice information, we apply the numeric variance algorithm~\cite{mishradata} for masking but preserving the date distribution.
Through verification by our collaborators, the data masking solutions provide adequate confidentiality to the raw records and prevent others from obtaining the actual identifications of taxpayers from the masked data.

\textbf{\jason{Network Construction.}}
\jason{Similar to prior research~\cite{tian_mining_2016}, we also fuse multiple data sources and construct a taxpayer network to facilitate further analysis. 
We first formulate the taxpayer network with taxpayers and investors, where the nodes can be both taxpayers and investors, and edges represent investment relationships.}
Then we provide more information about the nodes by fusing their profile information and audit records to the network.
A connected component in the resulting directed graph reflects that of \textit{a related party}.
We further adjust the taxpayer network according to our design requirements \textbf{(R1)} and domain knowledge.

Pruning is necessary as the original dataset contains almost 5 million nodes (i.e., taxpayers and investors), which creates a massive burden on the performance of the subsequent network fusion. 
We prune the related parties that contain only one taxpayer, which are not adequate enough for performing any related party transactions.
Our collaborators point out that according to the regulations, a necessary condition for convicting related party transactions is that there exists a common shareholder who holds 10\% or more of the voting rights.
We model this regulation by removing investors who have less than 10\% final investment ratio over any taxpayers from our network. The final investment ratio of an investor is defined as the maximum product of investment ratios for all paths starting from the investor to any end nodes.
\jason{We build a trade network with the invoice data filtered by the nodes in the taxpayer network, as taxpayers not in the network must be innocent of RPTTE. The trade network serves two purposes: it helps calculate the taxpayers' cumulative daily profits by aggregating all edges, and can be fused with the taxpayer network to obtain all related party transactions. If both the buyer and seller of an invoice belong to the same component in the taxpayer network(i.e., related party), it is \textit{a related party transaction}.}

\subsection{Data Analysis Module}
\jason{This module detects suspicious RPTTE groups by fusing the networks based on the parameters, and extracts features of each group to help further ranking and analysis.}

\textbf{Suspicious RPTTE Group Detection.}
{\name} provides interactive queries to assist tax administration officers in detecting any suspicious RPTTE groups within the period of concern and the ideal information-seeking space. 
\jason{Suspicious RPTTE group detection primarily performs network fusion, which aims to corroborate the suspiciousness of the related parties in the taxpayer network with invoices as evidence.
The edges in the trade network (i.e., transactions) are fused into the taxpayer network if both ends of the edge (i.e., buyers and sellers) exist and are reachable in the taxpayer network (i.e., related party).}
The resulting network is a directed graph of multiple disjoint components, i.e., suspicious RPTTE groups. 
We remove the groups that have not committed any related party transactions to present the most appropriate output.

On some occasions, such a method for detecting suspicious RPTTE groups generates an enormous related party due to a few super-connectors who have invested in various large corporations. 
We introduce two relevant thresholds to cut the large components and divide the huge groups, namely, \textit{the maximum transaction chain length} and \textit{the maximum control chain length}. 
\jason{Our collaborator tells us that tax evaders seldom commit RPTTE with loosely related taxpayers as it would involve more witnesses and thus increases the risk of being exposed.}
The maximum transaction chain length lets users decide to what extent the network fusion algorithm should connect the end nodes of any component with related party transactions.
It is defined as the maximum shortest path length for the buyer and seller to be considered reachable. 
Moreover, from the experience of domain experts, a related party involving historical tax evaders tends to commit tax evasion again. 
Therefore, nodes that have not conducted related party transactions should not be naively pruned as they might be former tax evaders. 
The maximum control chain length lets users decide the maximum shortest path length for investors and taxpayers to be considered reachable, which can effectively control the size of the related parties.

\textbf{Feature Extraction.}
\jason{We extract features that characterize the RPTTE group and the taxpayers (e.g., the number of taxpayers with evasion records and the cumulative daily profit of the taxpayers).
The cumulative daily profit modeled by accumulating the cash flows applies to invoices from the beginning to the end of a tax period.
We also extract features that delineate the related party transactions (e.g., the daily related party transaction amount and the number of effective related party transactions).}
We define a related party transaction as effective when the transaction takes place between two affiliates who differ in profit status, reported as one of the standard settings for transfer pricing~\cite{UN_manual_2017}. Clarifying, when buyers in profit purchase goods and services from sellers in loss, the tax evasion incentive is obvious and should be further investigated.
All the features extracted by the Data Analysis module are shown in the Visual Analysis module to help tax experts select, compare, and explore according to their priorities.
%\yong{I do not quite understand it here. Please refer to my chinese comments.}

\subsection{Visual Analysis Module}
The Visual Analysis module supports interactively mining and exploration of suspicious RPTTE groups through four coordinated views.
As shown in Fig.~\ref{fig:system_overview}, the interface of {\name} consists of four major UI components: 
\jason{a) The Control Panel configures the period and relevant thresholds for network fusion. 
b) The Group Overview displays any of the suspicious RPTTE groups ranked by selected features. 
c) The Graph View visualizes the ownership topology and trade relationships in the selected RPTTE group. 
d) The Detail View shows the cumulative profit status of the two taxpayers connected by the selected related party transactions. 
All the views are developed to bring forward different contexts for identifying suspicious RPTTE groups interactively.}

\subsubsection{Control Panel}
The Control Panel (Fig.~\ref{fig:system_overview}(A)) consists of a bar chart and a parameter selector to support network fusion through interactively setting the preferred period or relevant thresholds.
The bar chart on the right-hand side of the Control Panel offers a temporal summary of the daily related party transaction amount, which reveals a cyclic pattern where most of the peaks are near the end of the month. 
\jason{Since the profit manipulation should happen after tax evaders know how much taxable income they have, we speculate that this information can work as a visual cue to guide the tax administration officers in conducting their exploration of the suspicious RPTTE groups.}
Users can select their period of concern by brushing or clicking a bar to automatically select the quarter, which is a typical tax period (\textbf{R1}).

The parameter selector allows users to configure the maximum transaction chain length and maximum control chain length for the algorithm to determine the extent of the suspicious RPTTE groups (\textbf{R1}). 
\jason{As every combination of the period and relevant thresholds results in a different set of suspicious groups, a query by using taxpayer ID is available to track a particular taxpayer in different periods and locate it to see if the relevant thresholds have over-pruned or under-pruned the group.}
All the visualizations intend to help users narrow down or extend the information-seeking space on demand. 
\jason{The results of the algorithm show suspicious groups detected in the group overview.}

\subsubsection{Group Overview}
\begin{figure}[tb]
    \setlength{\belowcaptionskip}{-0.3cm}
	\centering
	\includegraphics[width=\columnwidth]{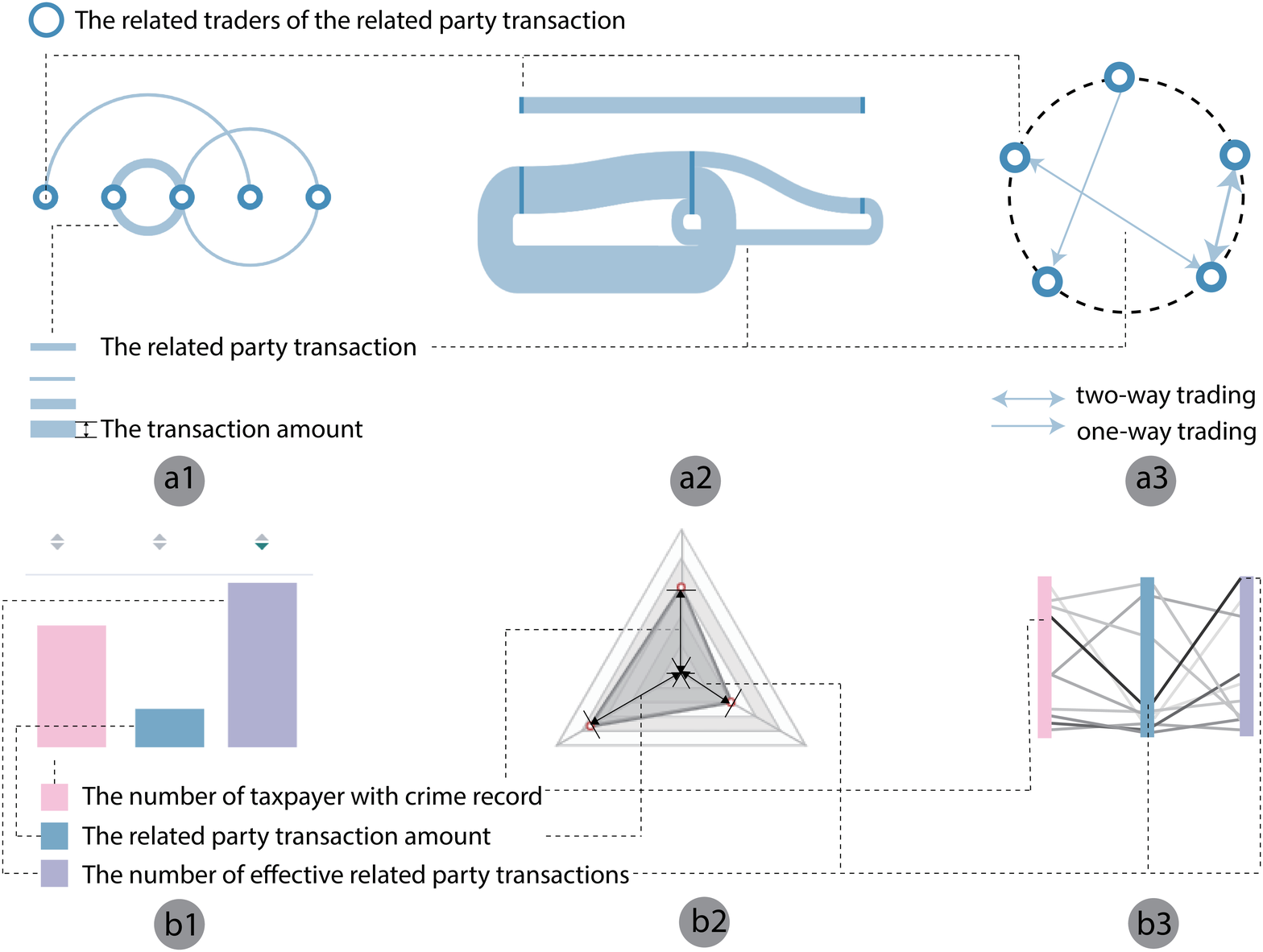}
    \caption{Alternative designs and visual encoding for visualizing the related party transactions in a group: (a1) arc diagram, \jason{(a2) Sankey diagram, (a3) chord diagram, and the legend for features: (b1) bar chart, (b2) radar chart, (b3) parallel coordinates plot. Our design used (a1) and (b1) in consideration of the design requirements.}}
    \label{group_view}
\end{figure}

\jason{The Group Overview (Fig.~\ref{fig:system_overview}(B)) shows a list of suspicious RPTTE groups, in which each row represents a suspicious RPTTE group, and consists of an arc-diagram based glyph
% \yong{Actually, is it better to call it ``node-link diagram'' instead of ``well-designed glyph''? Yating, please check if you would replace this term throughout the whole draft.}
and a bar chart to help users focus on the most suspicious groups (\textbf{R2}). 
We extend the arc diagram as a glyph to visualize the topology of the related party transactions, which allows users to estimate the complexity of the suspicious RPTTE group quickly.}
The nodes representing traders are arranged on a horizontal baseline, while the arcs indicating the related party transactions are drawn in a clockwise direction to show the trading direction. 
The profit flows from left to right are positioned above the horizontal line, and the flows from right to left are shown below to avoid visual clutter. 
The arc width encodes the transaction amount, which is also employed in the graph view, complying with the consistent encoding design principle~\cite{cao2011dicon}.

The bar chart shows three group features, including the number of taxpayers with tax evasion records, the related party transaction amount, and the number of effective related party transactions (\textbf{R2}).
\jason{By default, the number of effective related party transactions is selected to rank the groups because the feature is engineered to represent one of the standard settings for transfer pricing~\cite{UN_manual_2017}}. 
%\yong{I do not understand the reason here? Why the number of effective related parties? Better to shown in the detail view? It is not convincing.}
Users can click the sort icons located at the top of the list to sort the groups. 
\jason{Hovering over the nodes in the glyph highlights the corresponding node in the graph view.
In addition, clicking the row propagates the details of any suspicious RPTTE group to the graph view for further analysis.}

\textbf{Alternative Design.}
We have come up with several alternative designs to show the topology of related party transactions and the group features.
% Sankey diagram
For visualizing the topology of related party transactions, the Sankey diagram (Fig.~\ref{group_view}(a2)) is the first alternative design as it can intuitively show the transaction flows. 
\jason{However, it performed worse on two-way trading, which would result in severe overlapping between transaction flows. 
While it is possible to overcome the drawbacks by aggregating the flow of mutual transactions, the solution will compromise data accuracy by omitting the trade direction.}
%\yong{what are you talking about by the last sentence?}
% chord chart
Next, inspired by the chord diagram showing the connections between entities, we have designed the second alternative (Fig.~\ref{group_view}(a3)), which aligns the traders on a circle and uses double arrows to represent the trade direction. However, the problem is that the profit flow information would be lost.
\jason{To provide a general preview of the profit flow without compromising data accuracy, we design the glyph described above to show two-way profit flow, as shown in Fig.~\ref{group_view}(a1).}

There are three alternative designs to show the group features,
% \yong{for the three alternative designs, please refer to the figures instead of only b1, b2, b3.}
including \jason{bar chart (Fig.~\ref{group_view}($b_{1}$)), radar chart (Fig.~\ref{group_view}($b_{2}$)), and parallel coordinates plot (Fig.~\ref{group_view}($b_{3}$)).
% radar chart
The radar chart is useful in comparing multiple quantitative variables. 
However, given the large number of suspicious RPTTE groups returned by the algorithm, it is challenging to perform inter-comparisons. 
Sorting the groups with a radar chart is also non-trivial.
% PCP
Another design is the parallel coordinates plot. 
In contrast to the radar chart, which is used for one group, the parallel coordinates plot provides an overview of the global distribution of all groups' features. 
It also intuitively ranks and selects any outstanding groups by the position of their features. 
Unfortunately, information overload is a serious problem. The visual clutter, especially in the densely populated area, prevents users from accessing all cases, which is unacceptable.
Finally, we adopt a simple yet elegant bar chart (Fig.~\ref{group_view}($b_{1}$)) to show the features and provide a sort button to facilitate interactive exploration (\textbf{R2}).}

\subsubsection{Graph View}
\begin{figure}[tb]
	\centering
	\includegraphics[width=\columnwidth]{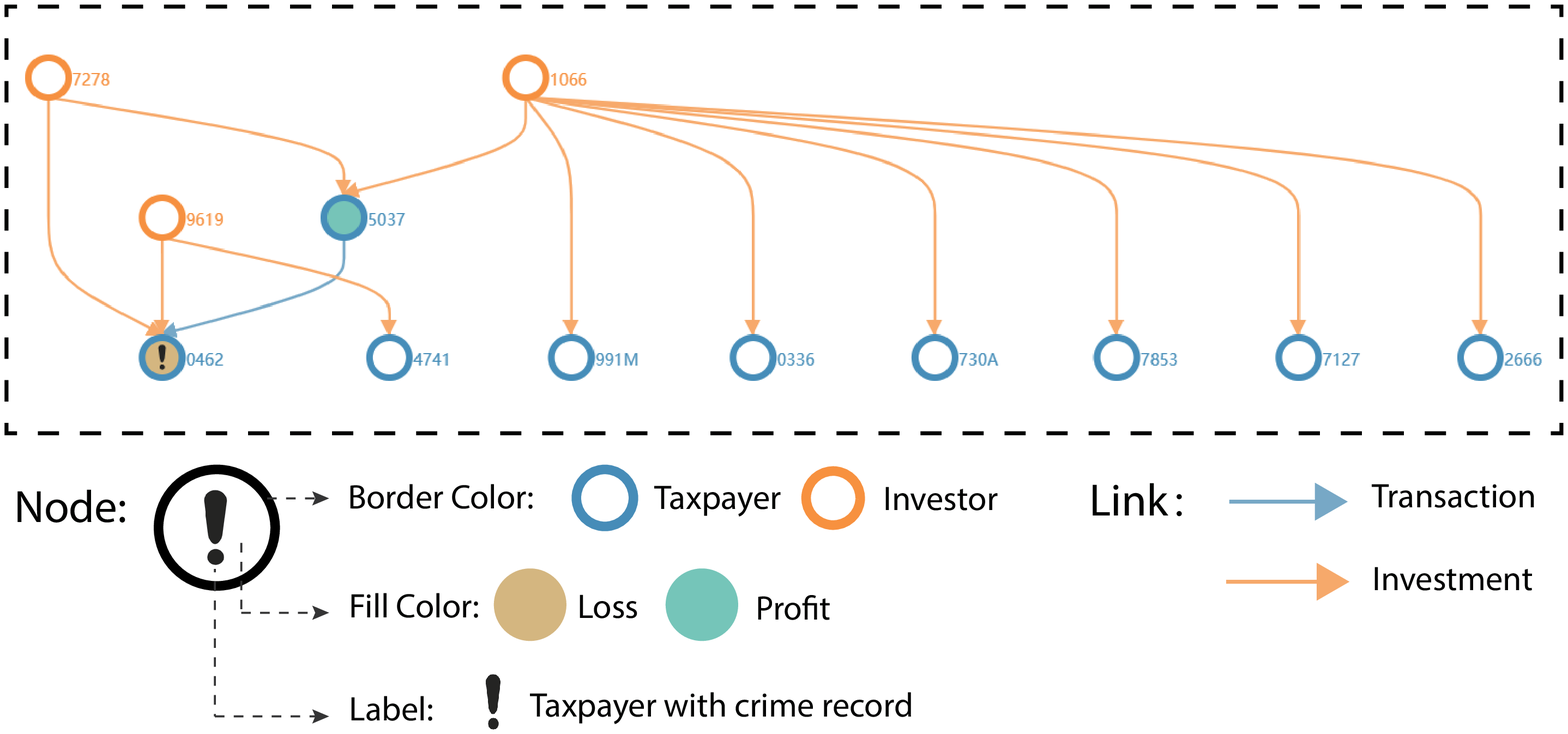}
    \caption{The visual encoding scheme for the Graph View.}
    \label{graph_view}
\end{figure}
\jason{The Graph View (Fig.~\ref{graph_view}) shows the hierarchical investment relationship and related party transactions within the selected suspicious group (\textbf{R3}). 
We employ the method proposed by Jünger and Mutzel~\cite{tamassia_2-layer_2002} to conduct the graph layout, as it is more efficient in a layered graph drawing and can effectively reduce the crossed links.}
%\yong{avoid mentioning the library in the method section.}
Also, we propose several encoding schemes to offer context for group assessment.
The color of the borders encodes the entity type of node (investor or taxpayer), and the corresponding period-end profit status for the node is encoded by the fill colors \jason{where the diverging color scheme is the same as the node color of the detailed view, as is introduced in Section~\ref{subsec-detail}}.
For the links, the type of relations uses color encoding. 
\jason{To stay consistent with the color encoding scheme of the node type, blue represents the related party transaction, while orange represents an investment.}
%\yong{trading vs. investment? pls confirm it.}
We emphasize the taxpayers with tax evasion records by drawing an exclamation mark in the circle.

\jason{The nodes are labeled with their IDs for users to track the entity down in the Control Panel.
Through the Graph View, users can gain an impression about whether the relevant thresholds have over-pruned or under-pruned the group and require a rerun of the network fusion algorithm with a different parameter setting (\textbf{R1}).
%\yong{what is the purpose of the ``if'' sentence here? Seems not relevant.}
By hovering over the related party transactions, the common beneficial owners and the entire ownership chain is highlighted to reveal deceptive cases where tax evaders use a complex ownership structure to hide their identities.
Clicking on the related party transaction link propagates the details to the detail view for further analysis.}

\textbf{Alternative Design.}
Alternatively, the force-directed layout offers more flexibility to allow users to explore the topology of the related party. 
\jason{However, it has been ruled out in the design because it confuses our target users with no clear and consistent orientations to follow for either the investors or the taxpayers.}
Another design was the Sankey diagram. However, it failed to handle multiple flows of different relationships for the related party.
Given the limitations of the previous two design choices, we applied the Dagre layout to accomplish our design requirements \jason{(\textbf{R3})}.

\subsubsection{Detail View}
\label{subsec-detail}

\begin{figure}[tb]
    \setlength{\belowcaptionskip}{-0.3cm}
	\centering
	\includegraphics[width = \columnwidth]{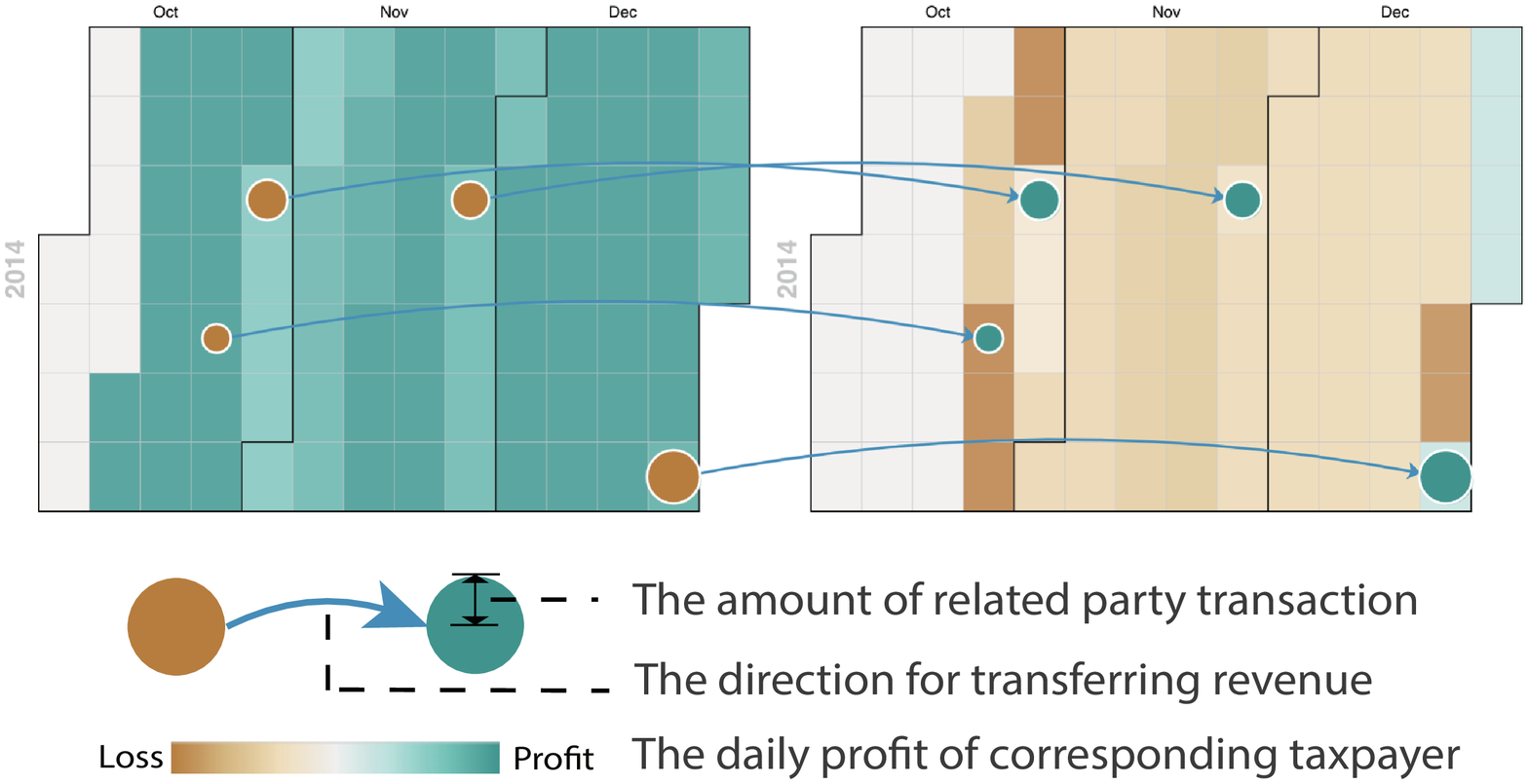}
    \caption{The Detail View shows the profit status of the traders who conducted the related party transaction, respectively.}
    \label{calendar_view}
\end{figure}
We proposed using two calendar heatmaps to visualize the profit status of two traders who conducted the selected related party transaction (\textbf{R4}). 
As shown in Fig.~\ref{calendar_view}, we encode the cumulative daily profit with the background color of each visual mark in the calendar heatmap.
\jason{We proposed two diverging color schemes with the help of ColorBrewer~\cite{ColorBrewer}, namely, the brown-blue-green and the red-yellow-green. 
The former is colorblind-friendly, and the latter is consistent with  conventional color usages in the Chinese stock market (i.e., red indicates profit while green represents loss).}

For each related party transaction, a dot is placed on the calendar heatmap to indicate the date, while its size encodes the transaction amount. 
The fill color of the dot indicates whether the transaction inflicts a profit or loss on the taxpayer. 
Each arrow connecting the dots indicates that the transaction takes place between the two taxpayers as well as the profit flow direction.
\jason{In most cases, the arrows are redundant because users can read from the fill color of the dots to determine the direction.
However, the transactions may be performed with other taxpayers in the same related party, so that the arrow indicators are essential to demonstrate the direction.
Nonetheless, we provide a button to help users remove them if severe visual clutter occurs.}
By comparing the colors of the visual marks on the same day, users can observe and assess the influence of the related party transaction on the profit status of the two taxpayers. 
A tooltip will appear to show the actual values and details of the invoice if users hover over a day in the calendar heatmap.
\jason{The toolbar supports rich interactions where users can toggle between the quarterly statements and yearly statements, or switch to another period to check their financial and transnational histories.}

\begin{figure}[tb]
    \setlength{\belowcaptionskip}{-0.3cm}
	\centering
	\includegraphics[width=\columnwidth]{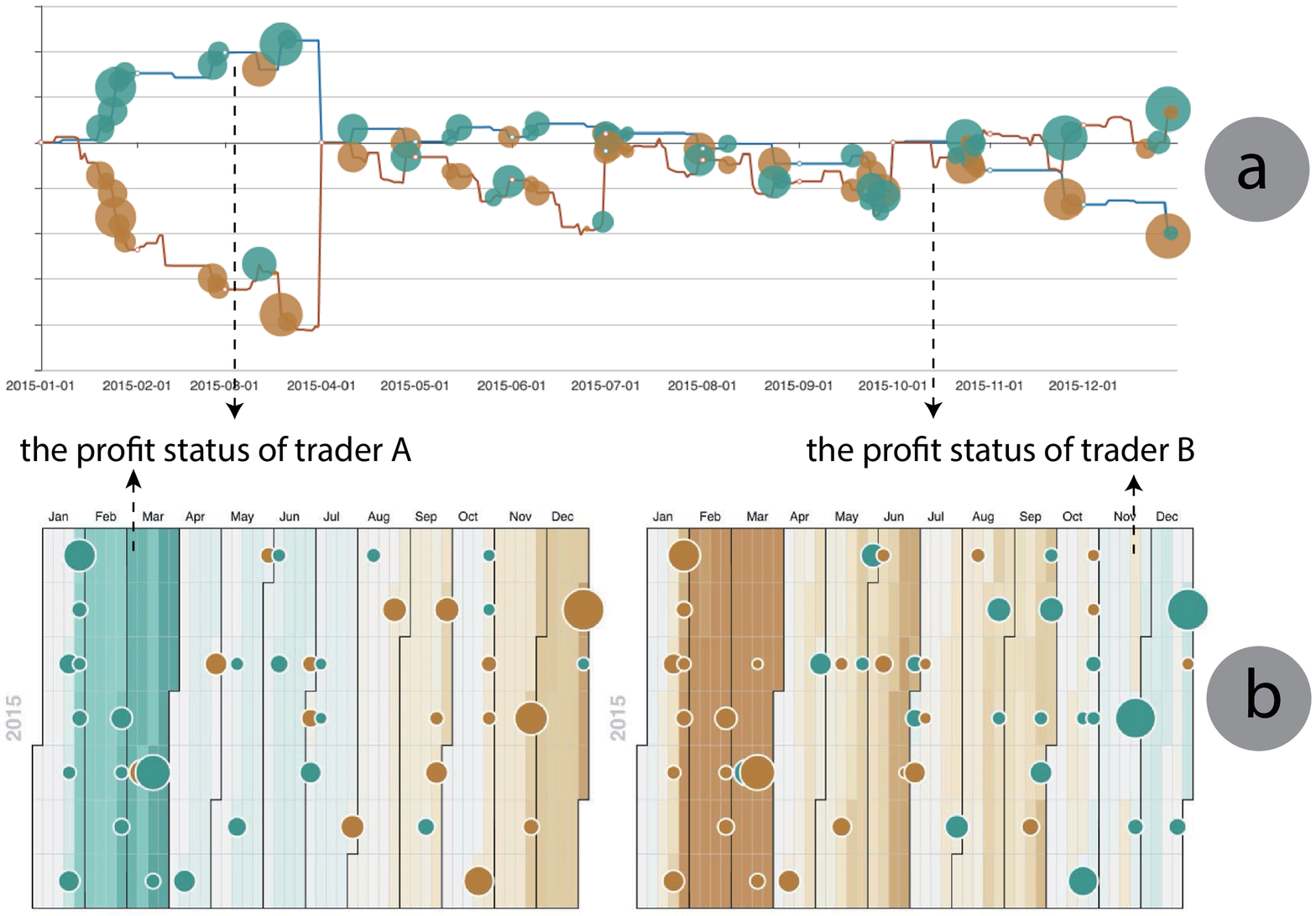}
    \caption{\jason{Design alternatives for visualizing the profit status of the traders who conducted the selected related party transaction: a) line chart, b) calendar heatmaps.}}
    \label{calendar_view_alternatives}
\end{figure}

\textbf{Alternative Design.}
\jason{To visualize the cumulative daily profit of the traders, we can replace the calendar heatmaps with a line chart, as in Fig.~\ref{calendar_view_alternatives}. 
The x-axis represents the date, and the y-axis represents the value of the cumulative daily profit.
Without radically changing the rest of the encodings, the line chart has the advantage of showing more clarity and scalability on some occasions.}

\jason{However, the line chart has a few shortcomings compared to the calendar heatmap.
First, since the profit status of taxpayers can fluctuate massively (especially for small and medium-sized enterprises), it can result in significant visual clutter when the two lines and x-axis intersect.
Second, the line chart can create an inconsistent visual impression for the same taxpayer. Due to the power difference between traders, the scale of the y-axis has to be adjusted accordingly to fit the limited design space. When users examine different related party transactions that involve the same taxpayer, they might find the same line being squeezed or stretched, creating difficulties for identification. 
Third, the line chart can be overly granular because inspectors care about the act of evasion more than the actual amount. For example, in identifying the effective related party transactions, users have to combine both visual channels: y-location of the line and color of the dots, to look for buyers in profit and sellers in loss. This combination is less intuitive than comparing the diverging colors of a rectangle and a dot in proximity in the calendar heatmap. We believe that the shortcomings of the line chart outweigh the benefits, leading to our final design choice.}
\section{Case Studies}
In this section, we present two real cases to demonstrate how {\name} can facilitate the detection and exploration of any suspicious RPTTE groups. 
The two cases were found by the experts when they used {\name} to explore tax evasion groups during our expert interviews, as will be introduced in Section~\ref{sec-expert-interview}.

\subsection{\jason{Beyond Financial Indicators: Catching Crafty Tax Evaders}}
\label{sec-case1}
E2 is interested in the integration of automated methods and visual analytics.
He was asked to evaluate the usefulness of {\name}. 
Without prior knowledge about the dataset, he first observed the bar chart in the control panel (Fig.~\ref{fig:system_overview}(A)) and noticed a cyclic pattern in the related party transaction amount. 

\jason{The patterns immediately drew his attention, and he selected the first quarter of 2014 by clicking on the highest bar in March.}
He did not change other default parameters for the algorithm because he did not know their actual impact \textbf{(R1)}. 
After running the network fusion algorithm, the suspicious RPTTE groups appeared in the Group Overview (Fig.~\ref{fig:system_overview}(B)).
% By the default descending order in the number of effective related party transactions, 

\jason{By sorting the suspicious groups in the descending order in terms of effective related party transactions, 
E2 noticed that the first group was very complicated as there were six taxpayers involved (Fig.~\ref{fig:system_overview}($b_{1}$)) \textbf{(R2)}. 
He was intimidated by its complexity and turned his attention to the second group, which had only two taxpayers committing related party transactions (Fig.~\ref{fig:system_overview}($b_{2}$)).}

E2 clicked on this group. Then the Graph View (Fig.~\ref{fig:system_overview}(C)) showed the topological relationships within the group (i.e., the investment and trading relationships) and the attributes of the investor and taxpayers (i.e., the taxpayer’s period-end profit status and historical tax evasion records) \textbf{(R3)}.
He first noticed that all blue nodes, which are the taxpayers, had an exclamation mark, which means they were all former tax evaders.

\jason{Then he observed that there was only one investor ($c_{1}$ in Fig.~\ref{fig:system_overview}(C)) who owned all the other companies, 
% judging by the border color of the nodes and the topology.
which is represented by the border color of the nodes and the topology.
He had identified the common beneficial owner with ease.}
Then, he turned his eyes to the fill color of the two nodes in the left and knew that there was a taxpayer at a profit ($c_{2}$ in Fig.~\ref{fig:system_overview}(C)) buying products from a taxpayer at a loss ($c_{3}$ in Fig.~\ref{fig:system_overview}(C)), indicating potential RPTTE behavior. 
The three visual cues guided him to pursue the details of the transaction, so he clicked on the blue edge connecting $c_{2}$ and $c_{3}$ to see how the related party transaction affected their cumulative daily profit.

The Detail View (Fig.~\ref{fig:system_overview}(D)) showed the profit status of the trades in the calendar heatmaps and the details of the transactions with dots \textbf{(R4)}. 
From the calendar heatmaps, E2 concluded that $c_{2}$ was always making money (Fig.~\ref{fig:system_overview}($d_{1}$)), while $c_{3}$ was gradually losing money (Fig.~\ref{fig:system_overview}($d_{2}$)). 
He recognized that all related party transactions were one-sided and inflicting cash flows from $c_{2}$ towards $c_{3}$. 
\jason{He also hovered over a few large dots to read the tooltip that shows the actual transaction amount. 
He suspected that the two related taxpayers could be repeated offenders of tax evasion.}
Then he navigated to other quarters and found similar trading patterns between 2014 and 2015. 
Through exploring different contexts of the suspicious groups, he believed that the suspicious group was an RPTTE group that should be further investigated. 
\jason{He pointed out that this group would likely be overlooked by traditional tax inspection methods using financial indicators, because its financial positions were stable and consistent.
He thought {\name} enabled the convenient exploration of useful context for the suspicious group and provides users with deep insights into catching crafty tax evaders, which may be difficult to achieve by using traditional methods such as checking financial indicators.
% to support convenient verification of the suspicious groups by automated methods and simplified the detection procedure.
}

\begin{figure}[tb]
    \setlength{\belowcaptionskip}{-0.3cm}
    \centering
    \includegraphics[width=\columnwidth]{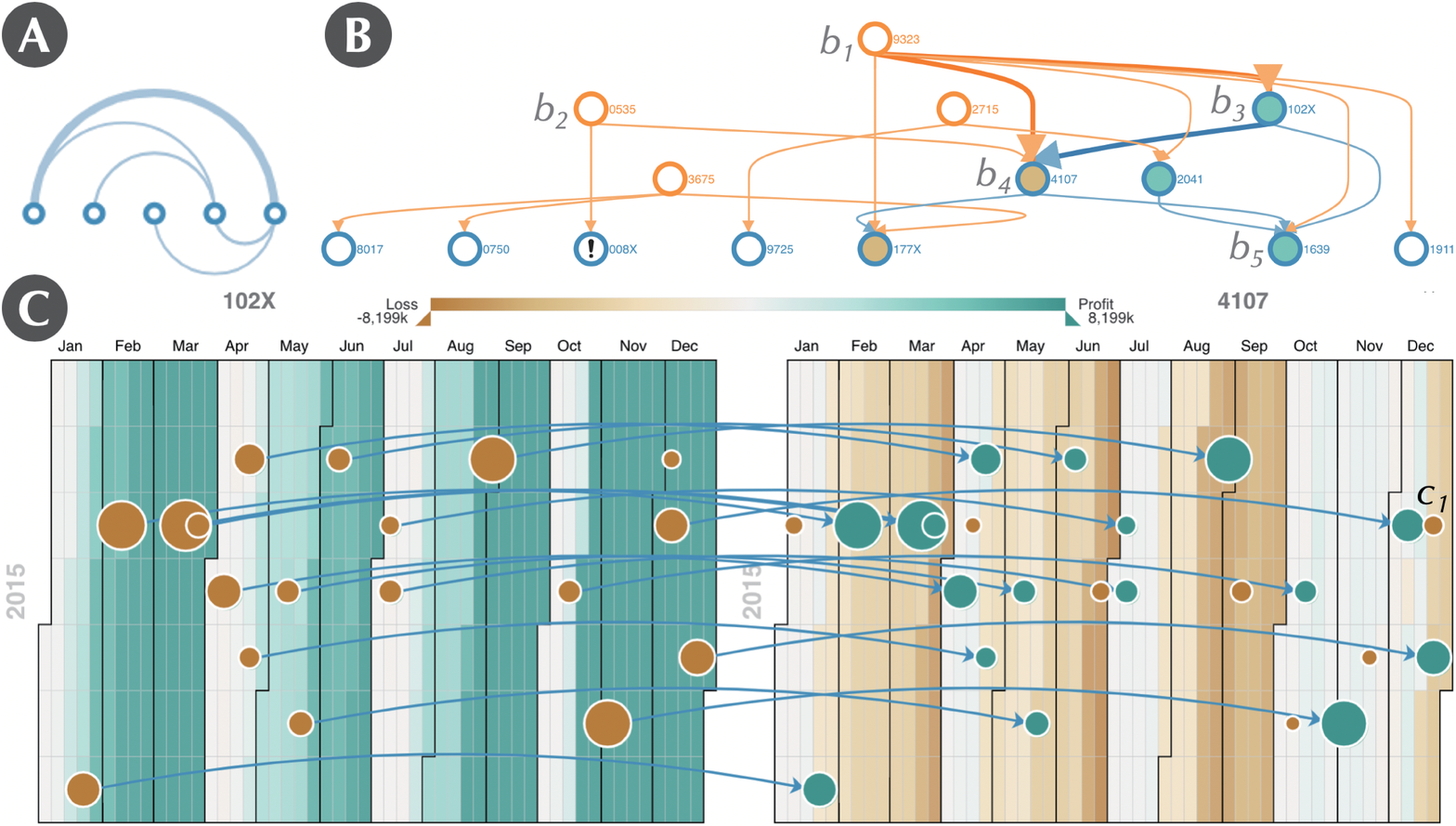}
    \caption{
       \jason{The expert E1 examined a RPTTE group with the help of three views: (A)The Graph Overview helped E1 find a group with complex trading relationships. (B)The Graph View showed that all taxpayers who conducted related party transactions had a common beneficial owner. (C)The left half of the Detail View showed several brown dots occurred in the calendar heatmap that is blue in general, indicating that the related party transactions inflicted an outflow when the taxpayer is in profit. Also, the arrows are pointing towards the right, indicating that revenue was transferring from a taxpayer in profit to a taxpayer in loss.}
      }
    %   \RongC{Please refer to visual cue about the findings.
    \label{case2}
\end{figure}

\subsection{\jason{Towards Deep Tax Inspection: Untangling the Tangled Transactions among Related Parties}}
E1 is interested in improving the audit process by effectively identifying the tax evasions.
He was asked to evaluate the usability of {\name}.
\jason{Being an experienced tax inspector, he decided to explore groups that were more complicated and had entangled trading relationships. 
He selected the second year (2015) as the period of interest and increased the maximum transaction chain length to allow more related party transactions to be included \textbf{(R1)}. }
The Group Overview then received from the network fusion algorithm and showed the suspicious RPTTE groups.

E1 ranked the groups by the related party transaction amount to search for groups with higher impact \textbf{(R2)}. 
Similar to Section~\ref{sec-case1}, after examining some groups with simple topological structures, he found a group with five taxpayers conducting related party transactions (Fig.~\ref{case2}(A)). 
He further checked the Graph View (Fig.~\ref{case2}(B)) and discovered that taxpayers $b_{3}$ and $b_{4}$ who are different in profit status, and the transaction was flowing from the taxpayer in profit (i.e., $b_{3}$) to the taxpayer in loss (i.e., $b_{4}$). 
He hovered over the transaction edge and noticed that the control chain was highlighted to indicate the common beneficial owner (i.e., $b_{1}$) \textbf{(R3)}.
\jason{He recognized that $b_{1}$ had control over all the taxpayers conducting related party transactions. 
Then he noticed that $b_{2}$, an investor of $b_{4}$, also invested in another company with historical tax evasion records.
He thought that $b_{4}$ was the most suspicious because $b_{2}$ may provide tax evasion instructions for $b_{1}$ and profit from it.
Based on the visual cues regarding the transaction direction and period-end profit status, he clicked on the trading relationship between $b_{3}$ and $b_{4}$ to conduct a deeper inspection.}

\jason{E1 had chosen the annual mode for calendar heatmaps (Fig.~\ref{case2}(C)) to display the profit status and all related party transactions in 2015 \textbf{(R4)}.}
He saw the arrows were pointing towards $b_{4}$, which indicate the direction of the transaction.
He found that all the transactions had been conducted when $b_{3}$ was in profit, and $b_{4}$ was in loss.
\jason{He noticed $c_{1}$ was not connected with an arrow, which he later found out that it was conducted by $b_{4}$ with $b_{5}$.
Although $b_{4}$ had trading relationships with three related taxpayers, E1 used {\name} to effectively untangle the complicated relationships and identify the most suspicious one with reasonable evidence.
After the exploration of the different contexts of the group, he believed that the group was very likely to be an RPTTE one.
He would further investigate $b_{1}$ and $b_{2}$ because they were the most suspicious taxpayers who conducted tax evasions in the group.
% He praised {\name} for providing audit directions in a complicated case.
E1 commented that he really enjoyed the power of {\name} in helping tax officers untangle complex transactions and providing guidance on finding highly suspicious taxpayers for deeper tax inspection.
}
\section{Expert Interview}
\label{sec-expert-interview}
% point out some questions and prepare some answers.
% 
\jason{We conducted remote interview sessions to verify the usefulness and effectiveness of {\name}.}
\yating{Each interview section lasted 60 minutes, with a 20-minute introduction to the system and task description through sharing our screen,} \jason{followed by} 30 minutes on the mining and exploration of suspicious groups by the experts themselves using our system with remote assistance, and finally 10 minutes for commenting on our system. 
The participants were experts \textbf{E1} and \textbf{E2}, who had been our collaborators since the beginning stages of our study. 
They helped with the formulation of design requirements, gave feedback to our design iterations, and evaluated our system. 

\yating{At the start of the interview sessions, we first introduced the four major views in {\name}, the functions, the visual encoding, and the interaction of each view, which aimed to fulfill the requirements they helped formulate in Section~\ref{sec-requirment}.
Then we demonstrated the system pipeline and used a simple case to demonstrate how to use our system to mine and explore suspicious RPTTE groups.}
\yating{After that, we provided remote control to enable them to use our system to identify suspicious tax-evasion groups on a real-world tax dataset.}
\jason{The experts had different focuses for the use of our system.
E1 was concerned with the usability of {\name} as to whether it could improve the audit process by identifying the tax evasions effectively, while E2 was concerned with the usefulness as to whether the integration of automated methods and visual analytics could improve the efficiency of the tax inspection process.}
\yating{After conducting the two interview sessions, we restructured their feedback into the following three perspectives to better summarize their opinions.}

\textbf{Profit Analysis Approach.} The experts agreed that the profit-oriented approach had the potential to discover RPTTE groups, which were often omitted in the current inspection routine. They recalled from their experience that such identification often relied on certain confidential financial indicators. Following accounting principles, these indicators are \jason{calculated from }period-end snapshots of the companies' financial \jason{positions, which are prone to deliberate profit manipulation near the end of a financial period}. Our system has provided a new perspective for taking the cash flow concept into consideration. Although E1 was skeptical about the practical value of monitoring profit by days and using it alone in the whole inspection routine, \jason{he} agreed that it was a good screening measure. E2 appreciated this approach because the \jason{profit and loss} information granted the tax officers more understanding of the continuous performance of the suspects. 

\textbf{System \jason{and Improvements.}} Both experts rated the system as useful and efficient for related transaction identification. They thought existing tools fell short before {\name} because of their inability to provide concrete evidence \jason{or} involve domain experts in the decision process. Besides that, the \jason{anomalies caught by the financial indicators could be a result of various factors like the macro-economic situation and seasonal industry patterns, rather than tax evasion. The }current methods generated a lot of false positives that required many inspection resources to verify. 

In comparison, E2 commented that\textit{''This system can save us much time in eliminating false positives ...It is intuitive to see the change in colors and how the taxpayer is affected by the transactions.''} E2 further suggested that the system could be improved by encoding extra tax-related information such as tax reduction policies for individual taxpayers. E1 \jason{gave credit for the entire analytics pipeline, especially showing the financial performance as context and the related party transactions as evidence. E1 advised} that the system could increase the decision ability if the unit price of the commodities could be incorporated. This is because, in the experts' experience, most convicted suspects used high-valued commodities to transfer profits. \jason{Additionally, the suspicion level increases significantly if the unit price of a transaction deviates vastly from the industry average. The advice is well-noted as a direction in the future iteration of the system.}

\textbf{Visualization and Interaction.} The experts found the visual encoding scheme intuitive and easy to learn. They understood the meaning behind the typical visual patterns. They valued the logical interactions between views as they allowed them to navigate and narrow down the search space quickly. Both experts stated that the Detail View was the systems' gem since it \textit{``decorated the plain cash-flow context with the significant impact of each related party transaction.``} E1 commented, \textit{``I had been dealing with numbers and tables for years ...This visualization is pleasing to the eyes.``} E2 added, \textit{``Sometimes the tax officers have different technical backgrounds. This system allows us to explain our audit justifications intuitively.``}
\section{Discussion}
While the case studies and expert interviews have corroborated the effectiveness and usefulness of {\name}, it is not without its limitations, and some design choices also need further clarification. 

\textbf{Scalability vs. Model Precision.} 
Scalability is an important aspect that we carefully considered during the development of {\name}. Due to the limited screen size, {\name} works well for suspicious tax evasion groups with less than 15 taxpayers. 
However, when there is a significantly large number of taxpayers in a suspicious tax evasion group, it would result in visual clutter in the \jason{Group Overview and Graph View}. 
According to our empirical observations, the majority of cases are often not \jason{as} extreme, and large corporate groups are scarce in reality. 
Also, due to their impact factor, large corporations are under close investigation in the current inspection routine; therefore, they are not the primary focus of the system. 
Users can change the related parameters in {\name} to shrink the size of suspicious groups \jason{at the cost of data} precision. 
\jason{In terms of visualization, a} possible improvement could be \jason{interactive clustering as it reduces the number of elements to be displayed without compromising data precision.
Besides, the dense graph can be broken into multiple views and apply focus–plus–context or overview-plus-detail techniques to allow} users to adjust the graph size dynamically \jason{and interactively}. 

\textbf{Data Availability vs. Design Choice.} 
The system relies heavily on the profit perspective to let users analyze related-party transactions. 
In the expert interview, the tax administration officer suggested that the unit price of commodities plays a significant role in tax inspection. 
In this study, the unit price is not available in the dataset, and the data granularity has limited the design choices in this study. 
However, once the unit price information is accessible, the information can provide more concrete evidence of how taxpayers conduct transfer pricing, which has been left for future work.
\jason{Without radically changing the current design, the unit price information can be encoded with the arrows in the Detail View, which is only an indication of transaction direction at the moment.
The redundancy in visual design has provided room for encoding newly available data.}

% \paragraph{Straightforward design vs. usability} 
\textbf{Straightforward Visualization Designs vs. Usability.} Overall, the visual designs of {\name} are straightforward and easy to understand. The major visualization designs of {\name} are also standard charts. For instance, the Group Overview of the suspicious RPTTE groups' features (Fig.~\ref{fig:system_overview}(B)) contains a bar chart, and the Graph View (Fig.~\ref{fig:system_overview}(D)) employs the popular visual design of node-link diagrams. These visualization designs are relatively simple. Such a design choice necessitates careful designs. The target users of {\name} are mainly tax inspectors with accounting backgrounds who often do not possess a deep understanding of visualization. Therefore, a visual analytics system with more standard charts or intuitive designs will be much easier for them to understand and explore, greatly enhancing the usability of {\name}.
\section{Conclusion and Future Work}

In this paper, we present {\name}, an interactive visual analytics system to help tax officers identify and examine suspicious tax evasion groups. The system integrates the traditional data mining algorithm and visual analytics techniques through an interactive analysis processing pipeline. We demonstrated the effectiveness and usefulness of {\name} through two case studies using real-world tax dataset and interviews with two domain experts.

In future work, we will consider acquiring the latest data sources that include the unit price in the invoice data. Also, the suspiciousness indicators can be customized to discover patterns that domain experts would like to explore, such as the tax burden difference and the rate of change in profit. Furthermore, the visual summary of our glyph design can be extended by incorporating the topological patterns in investment relationships as a widely used pattern matching method~\cite{tian_mining_2016}. Last, we would like to advance the usefulness of our system with an additional view to portrait groups with similar features and allow users to select similar cases if they find one suspicious case.

\acknowledgments{
We would like to thank the domain experts from SERVYOU group in China for the helpful discussions and the support of user studies.
This research was partially supported by “The Fundamental Theory and Applications of Big Data with Knowledge Engineering” under the National Key Research and Development Program of China with Grant No. 2016YFB1000903, the MOE Innovation Research Team No. IRT\_17R86, the National Science Foundation of China under Grant Nos. 61721002 and 61532015, HK RGC GRF grant 16213317 and Project of SERVYOU-XJTU Joint Innovation Center of Big Tax Data.}

\bibliographystyle{abbrv-doi}

\bibliography{ref}
\end{document}